\documentclass[prb,amsmath,amssymb,superscriptaddress,twocolumn]{revtex4}
\usepackage{braket}
\usepackage{graphicx}
\usepackage{bbm,color,ulem}
\DeclareMathAlphabet{\mathpzc}{OT1}{pzc}{m}{it} 
\begin{document}
\title{Studying many-body localization in exchange-coupled electron spin qubits using spin-spin correlations}
\author{Robert E.\ Throckmorton}
\author{S.\ Das Sarma}
\affiliation{Condensed Matter Theory Center and Joint Quantum Institute, Department of Physics, University of Maryland, College Park, Maryland 20742-4111 USA}
\date{\today}
\begin{abstract}
We show that many-body localization (MBL) effects can be observed in a finite chain of exchange-coupled spin qubits
in the presence of both exchange and magnetic noise, a system that has been experimentally realized in semiconductors
and is a potential solid-state quantum computing platform.  In addition to established measures of MBL, the level
spacing ratio and the entanglement entropy, we propose another quantity, the spin-spin correlation function, that can
be measured experimentally and is particularly well-suited to experiments in semiconductor-based electron spin qubit
systems.  We show that, in cases that the established measures detect as delocalized ``phases'', the spin-spin correlation
functions retain no memory of the system's initial state (i.e., the long-time value deviates significantly from the
initial value), but that they do retain memory in cases that the established measures detect as localized ``phases''.
We also discover an interesting counterintuitive result that there is no clear tendency towards localization with increasing
charge noise in small systems ($3$--$10$ spins).  The proposed experiments should be feasible in the existing semiconductor
spin qubit systems.
\end{abstract}
\maketitle

\section{Introduction}
One of the main platforms of current theoretical and experimental interest for quantum computation is the semiconductor-based electron spin
qubit.  The main advantages of this semiconductor platform are its fast gate operation time and  its eventual scalability, with the main
drawback being that the current state-of-the-art systems still have only $1$--$3$ operational qubits at most.  All such systems consist
of individual electrons confined to quantum dots that are defined by gate voltages.  Even though much progress has been made in improving
these systems, they continue to lag behind other platforms in terms of fidelity of single- and two-qubit gates.  While single-qubit gate
fidelities of $99.9\%$ have been reported\cite{VeldhorstNatNano2014}, two-qubit gate fidelities have only gone as high as $98\%$, which is
well below the error correction threshold\cite{HuangNature2019}.  These modest fidelities, unfortunately, were only achieved with the qubit(s)
being operated in isolation---no other qubits were present.  In these systems, the only sources of error are magnetic (Overhauser) noise
and electronic charge noise affecting the exchange couplings between electrons.  Experimental systems with more than two quantum dots also
exist\cite{GullansPRB2019,CerfontainePRB2020,ChanArXiv,GilbertArXiv}, but the coupling of multiple qubits beyond the qubit(s) being operated
on introduces crosstalk, which causes further degradation of fidelity.  The invariable presence of noise and the modest number of operational
qubits have hindered progress in spin qubits although its promise as a quantum computing platform, because of fast gate operations and
efficient scalability, remains viable.  An important question in this context is whether anything worthwhile can be done with a few spin qubits
beyond demonstrating their isolated gate operations. 

The presence of Overhauser and electronic noise, however, should enable the study of MBL effects in these systems, as they are analogous to
tight-binding chains with random on-site potentials and randomized tight-binding couplings.  In this sense the existing spin qubit systems enable
an almost ideal emulation of MBL in disordered Heisenberg spin chains, which would demonstrate the viability of spin qubits as a solid state
dynamical quantum simulator.  Such dynamical studies can be performed now in current experimental semiconductor-based electron spin qubit systems,
as one is only concerned in this case with measuring the natural evolution of the system and not with precise control of spins as in quantum
computation.  One simply needs to prepare the system in a given state, let it evolve naturally, and then measure the spins.  Since the decoherence
time is rather long in the already-existing spin qubit systems, carrying out a dynamical MBL study is timely and  could serve as an important
milestone in the progress of the subject.  In this context, we mention that so far the only quantum simulations successfully achieved in semiconductor
quantum dot spin qubit platforms are static ground state emulations of Mott-Hubbard physics\cite{HensgensNature2017,SaffordPRL1994,ButerakosPRB2019}
and quantum ferromagnetism\cite{ButerakosPRB2019,DehollainNature2020,WangPRB2019}.  Simulating MBL or even just Anderson localization, which has
so far only been achieved in atomic quantum gas systems\cite{KohlertPRL2019,LueschenPRL2018,SchreiberScience2015} would therefore be an important
experimental achievement for solid state spin qubit platforms.  In the current work, we point out a rather convenient way of studying MBL in spin
qubit systems which should be feasible in the existing few qubit semiconductor quantum dot platforms.

Ever since the work of Anderson\cite{AndersonPR1958} and especially of Basko, Aleiner, and Altshuler\cite{BaskoAnnPhys2006,BaskoPRB2007},
the topic of many-body localization has come to be of great interest in physics.  Even though the former discussed
the role of interactions in the localization physics described therein, the latter sparked great interest in the topic,
considering dynamical aspects including the  whole spectrum not limited just to the ground state localization question.
Since then, many works, both theoretical and experimental, have addressed the topic of many-body localization\cite{NandkishoreARCMP2015,SmithNatPhys2016,SchreiberScience2015,ChoiScience2016,OganesyanPRB2007,ZnidaricPRB2008,PalPRB2010,BardarsonPRL2012,SerbynPRL2013,HusePRB2014,SerbynPRX2015,LuitzPRB2015,IslamNature2015,BarnesPRB2016,VasseurJSM2016,GoihlQuantRep2019,OrellPRB2019,SunPRR2020,KieferEmmanouilidisPRL2020,GopalakrishnanPhysRep2020}.
Many-body localization is the survival of some remnant of Anderson localization in the presence of interactions.  In
particular, it relates directly to the question of whether a disordered interacting many body system can self-thermalize
when interparticle interactions are turned on even when the corresponding noninteracting system is localized.  While
this was originally considered in the context of particles in a tight-binding chain, such effects may also be explored
in a chain of exchange-coupled spins, which is mathematically equivalent via a Jordan-Wigner transformation.  It is this
system, which is the type of system realized by experimental semiconductor-based spin qubit systems, that we will focus
on in this work.

Semiconductor spin qubits are particularly well suited for studying MBL in a solid state system for several reasons: (1)
The interqubit exchange coupling can be controlled rather accurately; (2) the system Hamiltonian is well approximated by
a nearest-neighbor Heisenberg chain\cite{HuPRA2000,ScarolaPRA2005}, where disorder-tuned MBL in an interacting one-dimensional
system is well established\cite{PalPRB2010,LuitzPRB2015,BarnesPRB2016}; (3) the system is well isolated from the surrounding
thermal bath in the sense that the dissipative energy relaxation is very small, with very long spin relaxation times; (4) the
relevant disorder from the Overhauser magnetic noise of the nuclear spin bath can be controlled to a reasonable extent; (5)
the system Hamiltonian is dynamically well controlled due to the all-electrical gate operations; (6) various fast measurements
allow defining the spin states of the system rather accurately as a function of time; (7) individual spins can be initialized
suitably to create well-defined initial states; (8) final spin states after time evolution through various unitary operations
can be measured at various times, allowing a comparison between initial and final states. 

We note that MBL has not yet been reported in any experimental solid state spin qubit systems.  In fact, the experimental observations of
MBL in atomic systems all suffer from severe finite-time effects since eventually the system thermalizes either through spontaneous
emission or through environmental coupling.  It will therefore be quite useful to see limited signatures of MBL in the solid state
spin qubit systems, even if such an experiment is restricted only to a few qubits since it will point to the MBL issues specific to
solid state systems, which typically couple to the environment more strongly than atomic systems.  In addition, MBL requires a
decoupling from the environment (i.e., low decoherence), which is also a stringent requirement for quantum computing.  Thus, our
proposed MBL study in spin qubits should help the development of quantum computing capabilities of semiconductor spin qubit platforms.
Our work should be viewed more in the context of what useful physics can be done with only a few spin qubits rather than as a study
of MBL itself in great depth.

A number of measures have been proposed for detecting many-body localization.  The two standard theoretical measures are the
average level spacing ratio and the long-time entanglement entropy.  While they are of great theoretical value, these metrics
are difficult, if not impossible, to measure experimentally.  As a result, another metric, proposed in the context of spin
systems\cite{WeiPRL2018} related to the spin-spin correlation length and measurable via NMR techniques in nuclear spin systems,
was recently suggested.  We propose here yet another metric that is particularly well suited to MBL studies on semiconductor-based
quantum dot electron spin systems, the disorder-averaged spin-spin correlation function, $[\braket{S_{j,z}S_{k,z}}](t)$, where
$j$ and $k$ label the spatial positions of two electron spins in the chain.  The utility of this metric for such systems in particular
is due to the fact that one can measure individual spins and consequently directly obtain the spin-spin correlation functions.
We will be working with a dimensionless version of this measure, $[\braket{\sigma_{j,z}\sigma_{k,z}}](t)$, where $\sigma_{j,z}$
is the $\sigma_z$ Pauli matrix acting on spin $j$.  This consideration is motivated by the fact that localization affects the
spatial spread of the wave function of the electron spin over time.  We thus expect that the spin-spin correlation function should
deviate significantly from its initial value if the system is in a delocalized ``phase'', but should remain close to its starting value
in the localized ``phase''.  Experimentally, this quantity can be directly measured dynamically in a semiconductor quantum dot based
spin qubit platform, and therefore, the delocalized and MBL ``phases'' could be distinguished.  To provide an illustration, we will consider
the case in which we initialize the spin chain in the state in which the leftmost spin is up, but the rest are down, i.e.,
$\ket{\psi(0)}=\ket{\uparrow\downarrow\cdots\downarrow}$.  In this case, at $t=0$, the leftmost spin is perfectly anticorrelated with
the other spins, i.e., $\braket{\sigma_{1,z}\sigma_{k,z}}(t=0)=-1$ for all $k>1$.

We calculate the average level spacing ratio for chains of spins of lengths $L=3$ to $10$ (i.e., a spin chain containing $3$--$10$
localized electron spins in quantum dots arranged in a one-dimensional array with controllable interspin coupling) and the long-time
entanglement entropy and spin-spin correlation functions for chains of lengths $L=2$ to $8$ (it is impossible to define a level spacing
ratio for $L=2$ within the $S_z=1-\tfrac{1}{2}L$ subspace, as there are only two levels in that subspace).  We assume
quasistatic electronic and Overhauser noise, which appear as noise in the exchange couplings and the magnetic field, respectively.
The quasistatic noise approximation is valid here since the typical gate operations times are much faster than the time scale for
any dynamical noise fluctuations.  We assume that the exchange interactions in the presence of noise follow a Gaussian distribution
of mean $J_0$ and standard deviation $\sigma_J$, truncated to positive values of the exchange (because the exchange coupling cannot
change sign in quantum dot based systems), and that the magnetic field fluctuations, arising from background slow nuclear dynamics,
also follow a Gaussian distribution, with zero mean and standard deviation $\sigma_h$.  The Gaussian random noise approximation is
widely used for spin qubits and should be well valid given the unknown random nature of the background noise fluctuations.  For
the two metrics that depend on time, the entanglement entropy and the spin-spin correlation function, we determine both at the
time, $J_0t=100$.  Most of this work must be done numerically, though some analytic expressions can be obtained for $L=2$.  We
demonstrate that the results for the spin-spin correlation function indeed behave as we expect---we find that they stay close to
their initial values in cases that the level spacing ratios and entanglement entropy indicate localization, while they deviate
significantly in cases where the other two metrics would indicate a delocalized ``phase''.  We also note that the functions initially
oscillate, but are damped out over a time scale $T_S$.  We determine $T_S$ by inspecting the calculated spin-spin correlation
results; we find that $T_S$ decreases with increasing $\sigma_h$ for fixed $\sigma_J$ in the delocalized ``phase'' but becomes constant
in the localized ``phase''.  We believe that the currently existing semiconductor  spin qubit systems with only a few dots should
enable a direct observation of the predicted saturation of $T_S$, thus indicating MBL.

The rest of the paper is organized as follows.  In Sec.\ \ref{sec:Model}, we state the model that we use in our calculations.  We
review and calculate the standard MBL measures in Sec.\ \ref{sec:EstMBLMeasures}.  We then determine the spin-spin correlation
function in Sec.\ \ref{sec:SpSpCorr}.  Finally, we conclude in Sec.\ \ref{sec:Conc}.  We present results for the average level
spacing ratio in large systems ($L=50$--$200$) in Appendix \ref{app:AvgLSR_LargeSystems}, as well as results for smaller systems, but
determined from all energy levels, in Appendix \ref{app:AvgLSR_AllEnLevels}.

\section{Model} \label{sec:Model}
The model that we consider here is a one-dimensional array of quantum dots with localized electrons representing a chain of spins,
all spin $\tfrac{1}{2}$ as appropriate for semiconductor spin qubits, of length $L$ (with $L$ corresponding to the number of quantum
dots containing localized electrons) coupled by nearest-neighbor Heisenberg exchange couplings:
\begin{equation}
H=-\sum_{k=1}^{L-1}J_k\vec{\sigma}_k\cdot\vec{\sigma}_{k+1}+\sum_{k=1}^L h_k\sigma_{k,z}, \label{Eq:Hamiltonian}
\end{equation}
where $\vec{\sigma}_k$ is the vector of Pauli matrices acting on spin $k$.  Equation \eqref{Eq:Hamiltonian} represents the well-established
Hamiltonian for the semiconductor spin qubit array with exchange spin coupling controlled by tunnel barriers between the dots.  We assume that,
ideally, all exchange couplings are equal (i.e., all $J_k=J_0$) and that there is no magnetic field (i.e., all $h_k=0$). However, we assume,
as appropriate for the experimental situation, the presence of quasistatic electronic and Overhauser noise in the system, so that there will
be a $k$ dependence in $J_k$ and $h_k\neq 0$.  We consider $L$ from $2$ to $8$.  Throughout this work, we model the quasistatic noise as random
Gaussian distributions of $J_k$,
\begin{equation}
f_J(J)\propto e^{-(J-J_0)^2/2\sigma_J^2},
\end{equation}
where $J\in [0,\infty)$ (i.e., truncated so that the $J_k$ are always positive, as is the case in actual experimental spin qubit systems), and
of $h_k$,
\begin{equation}
f_h(h)\propto e^{-h^2/2\sigma_h^2}.
\end{equation}
We will also take $\sigma_J=0.1J_0$ in all calculations in which we vary $\sigma_h$.  Similarly, when we vary $\sigma_J$, we fix $\sigma_h=0.1J_0$.
These disorder choices are typical for semiconductor spin qubit systems, but changing these noise variances to other reasonable values do not change
any of our conclusions.

\section{Established MBL measures} \label{sec:EstMBLMeasures}
Before we present our results for the spin-spin correlation functions, we first review the standard
MBL measures, namely, level spacing ratios and the entanglement entropy, and present results for them in
order to more clearly demonstrate the utility of spin-spin correlations as an alternative MBL measure.
We emphasize that there is no possibility of ever measuring level spacing and entanglement entropy in
semiconductor spin qubits, but measuring the spin-spin correlator should be straightforward in existing
systems containing two or more qubits.  Thus, Secs.\ \ref{subsec:LevelSpacing} and \ref{subsec:EntanglementEntropy}
below are provided for completeness, and in order to connect with the standard MBL theoretical literature
in the context of spin qubit physics. Neither the level spacing statistics nor entanglement entropy is
ever directly measured in any MBL laboratory experiments.

\subsection{Level spacing ratios} \label{subsec:LevelSpacing}
We first consider the level spacing ratios, which are defined as
\begin{equation}
r^{(n)}_\alpha=\frac{\min[\delta^{(n)}_\alpha,\delta^{(n+1)}_\alpha]}{\max[\delta^{(n)}_\alpha,\delta^{(n+1)}_\alpha]}, \label{Eq:LSDef}
\end{equation}
where $\delta^{(n)}_\alpha=E^{(n+1)}_\alpha-E^{(n)}_\alpha$ is the gap between energy levels $n$ and $n+1$ (we assume that
the levels are sorted in ascending order, so that $E^{(n+1)}_\alpha\geq E^{(n)}_\alpha$) and $\alpha$ labels a given realization
of disorder.  We are in particular interested in the disorder and level average of this ratio, which is denoted by $[r^{(n)}_\alpha]$.
In the delocalized ``phase'', the level spacings follow a Gaussian orthogonal ensemble (GOE) distribution, yielding a disorder- and
level-averaged level spacing ratio of\cite{AtasPRL2013} $[r^{(n)}_\alpha]\approx 0.53$.  On the other hand, in the localized ``phase'',
the level spacings have a Poisson distribution, which gives\cite{AtasPRL2013} $[r^{(n)}_\alpha]=2\ln{2}-1\approx 0.39$.

\begin{figure}
	\centering
		\includegraphics[width=\columnwidth]{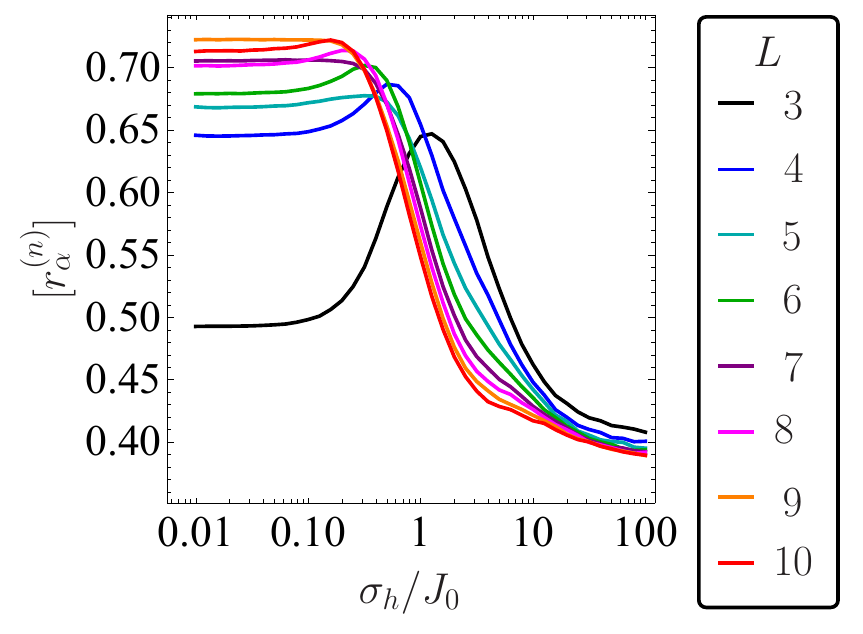}
	\caption{Disorder-averaged level spacing ratios $[r^{(n)}_\alpha]$ as a function of $\sigma_h/J_0$ for $L=3$
	to $10$ and $\sigma_J/J_0=0.1$, calculated only using levels in the $S_z=1-\tfrac{1}{2}L$ subspace.}
	\label{fig:rsAllLengthsPlot_1SUSSOnly}
\end{figure}

Because the integrals giving $[r^{(n)}_\alpha]$ cannot in general be done analytically in the presence of
random disorder, we evaluate this ratio numerically.  We use $50,000$ realizations of disorder in our calculation
and average over all energy levels within the $S_z=1-\tfrac{1}{2}L$ subspace.  We determine $[r^{(n)}_\alpha]$ as
a function of $\sigma_h/J_0$ and plot the results in Fig.\ \ref{fig:rsAllLengthsPlot_1SUSSOnly}.
We see that, with increasing Overhauser noise, there is a clear difference between the delocalized and MBL
``phases'' of this system for most lengths.  However, we see an interesting phenomenon in our results---while
the curves tend toward the Poisson value for large $\sigma_h$, we find that they exceed the GOE value for small
$\sigma_h$, going as high as $0.72$.  This is likely a finite-size effect, as we illustrate in the Appendix;
we find there that the values obtained for small $\sigma_h$ for large systems ($L=50$--$200$) decrease with
increasing system size.  Strictly speaking, one does not obtain true MBL physics except in very large systems
(say, $L=1,000$ or larger), but we see that, even at $L=10$, some signatures are already visible.  The crossover
between the delocalized and MBL ``phases'' happens for larger values of $\sigma_h$ for shorter chain lengths.
For example, we see a crossover between these two ``phases'' at $\sigma_h\approx 2J_0$ for $L=6$, but at
$\sigma_h\approx 1.2J_0$ for $L=8$.  Thus, it should be easier to induce MBL in systems with a larger number of
qubits, but it is interesting to note that the MBL effectively manifests itself in Fig.\ \ref{fig:rsAllLengthsPlot_1SUSSOnly}
even for $L=3$ although the corresponding delocalized phase is only apparent in Fig.\ \ref{fig:rsAllLengthsPlot_1SUSSOnly}
for $L>4$.  Since semiconductor spin qubit platforms with $2$--$5$ spins are already available, we believe
that our proposed MBL experiments can be carried out in semiconductor quantum dot arrays.

We note that, for $L=3$, $[r^{(n)}_\alpha]$ has a nonmonotonic behavior, with a peak around $\sigma_h\approx 1.9J_0$.
We also see a similar, but much less pronounced, peak for the other values of $L$ shown.  In general, the
systems that we are studying are small, and we do not expect $[r^{(n)}_\alpha]$
to tend precisely to the GOE or Poisson values for such system sizes due to the small number of energy
levels present.  But our reason for studying small systems is deliberate as we are interested in addressing
the question of doing some useful dynamical physics in the existing, and not hypothetical, spin qubit
systems, and the current semiconductor spin qubit platforms have $L=1$--$6$ coupled spins at most.
Therefore, the theoretical situation in the thermodynamic limit, or even for $L=20$, is not relevant
for our consideration.  We want to know what can be done with the existing spin qubit systems ($L<8$
for sure) in semiconductor quantum dots.

\begin{figure}
	\centering
		\includegraphics[width=\columnwidth]{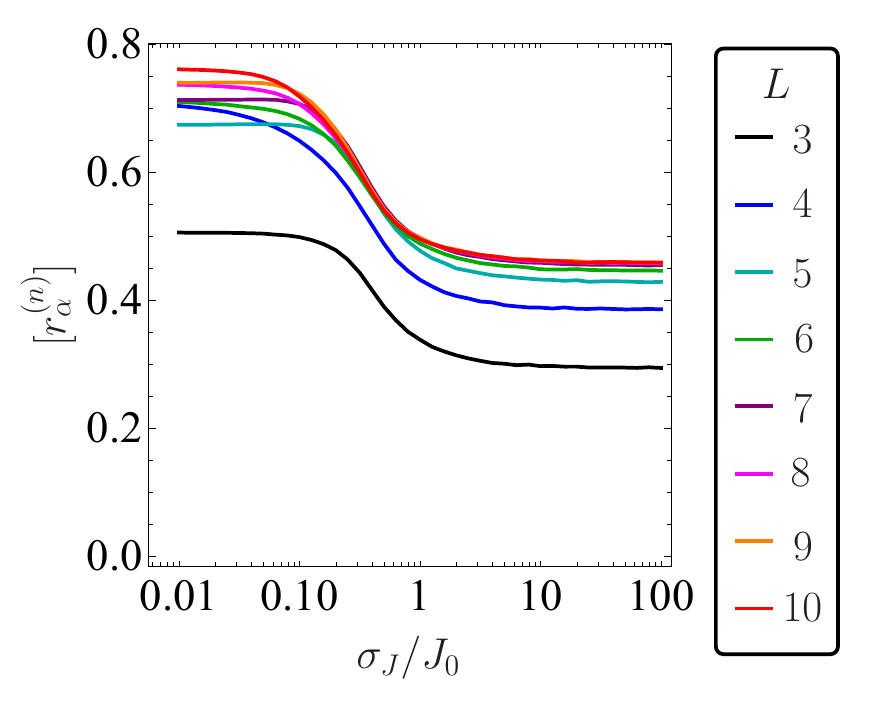}
	\caption{Disorder-averaged level spacing ratios $[r^{(n)}_\alpha]$ as a function of $\sigma_J/J_0$ for $L=3$
	to $10$ and $\sigma_h/J_0=0.1$, calculated only using levels in the $S_z=1-\tfrac{1}{2}L$ subspace.}
	\label{fig:rsAllLengthsPlot_varysJ_1SUSSOnly}
\end{figure}

We also do the same calculation, but with $\sigma_h$ fixed and varying $\sigma_J$; we present our results
in Fig.\ \ref{fig:rsAllLengthsPlot_varysJ_1SUSSOnly}.
Here, we find that $[r^{(n)}_\alpha]$ converges to specific values both for very small and very large $\sigma_J$,
but that these values are neither the GOE value for small $\sigma_J$ nor the Poisson value for large $\sigma_J$.
In fact, these results imply that there is never a clear tendency towards a localized ``phase'' for any system
size shown in the plot.  This is likely just a finite-size effect; as shown in Appendix \ref{app:AvgLSR_LargeSystems},
the large $\sigma_J$ value tends closer to the Poisson value for large $L$, while the small $\sigma_J$ limit decreases
with increasing $L$.  However, it does so more slowly than in the case of fixed $\sigma_J$ and varying $\sigma_h$.
This rather interesting, and somewhat surprising, result is specific to semiconductor spin qubits where charge
noise is capable of randomly enhancing the interaction effect by increasing the exchange coupling because of the coupling
being constrained to be positive.  This happens due to the fact that charge fluctuations could cause a random
suppression of the tunnel barrier between two dots, thus randomly increasing the wave function overlap, and
hence, the exchange coupling\cite{HuPRL2006,CulcerAPL2009}.  An experimental observation of the apparent persistence
of the delocalized ``phase'' in spin qubit systems with increasing charge noise will be an interesting counterintuitive
discovery.

We also performed the above calculations, but using all energy levels instead of just those within the $S_z=1-\tfrac{1}{2}L$
subspace, and present these results in Appendix \ref{app:AvgLSR_AllEnLevels} for the sake of comparison with the $S_z=1-\tfrac{1}{2}L$
subspace results.  We find that the same qualitative conclusions may be drawn from these results as well.

\subsection{Entanglement entropy} \label{subsec:EntanglementEntropy}
We next consider the entanglement entropy.  The long-time entanglement entropy is simply the long-time limit
of the von Neumann entropy of a subsystem $A$ of the full system $S$ after the environment $B$ (i.e., the parts
of $S$ not contained in $A$) is traced out:
\begin{equation}
S_E=-\mbox{Tr}_A\,\rho_A\ln{\rho_A}, \label{Eq:EEDef}
\end{equation}
where $\rho_A=\mbox{Tr}_B\,\rho$ is the density matrix of $A$ and $\rho$ is that of $S$.  This measure, unlike
the level spacing ratio, is time-dependent, but saturates to a constant value at long times in the presence of
disorder.  In the delocalized ``phase'', the entanglement entropy follows a volume law (i.e., it depends on the size
of the ``bulk'' of the system, in our case the number of spins $L$ in the chain), while it follows an area law
(i.e., it depends on the size of the ``surface'', which is just two points in our one-dimensional case) in the
localized ``phase''.

\begin{figure}
	\centering
		\includegraphics[width=0.49\columnwidth]{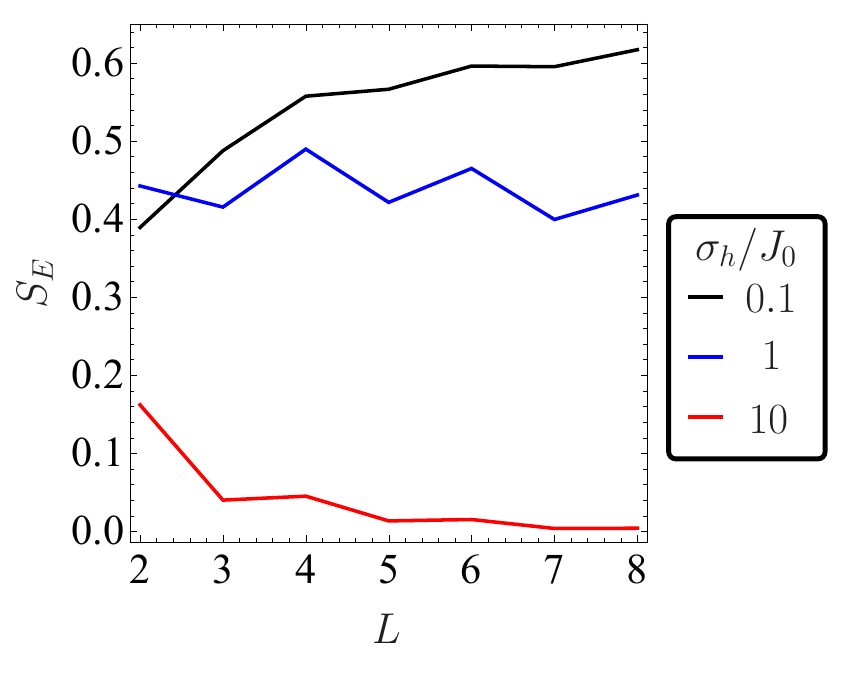}
		\includegraphics[width=0.49\columnwidth]{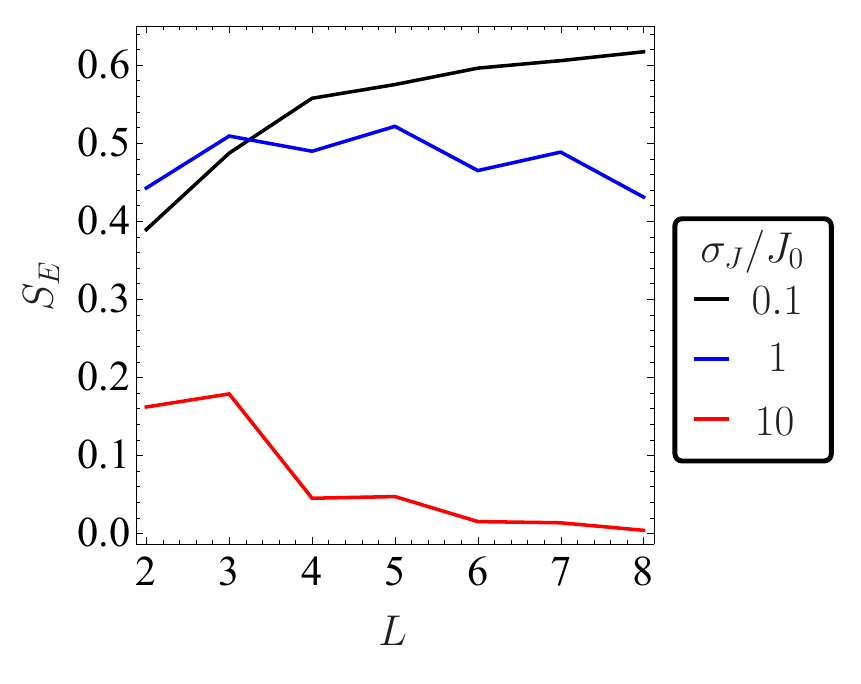}
	\caption{Plot of the entanglement entropy $S_E$ of the spin chain, with one half of the chain designated
	as the subsystem and the other as the ``environment'' as a function of spin chain length for $\sigma_h/J_0=0.1$,
	$1$, and $10$, at $J_0t=100$ and $\sigma_J/J_0=0.1$.  We assign the middle spin for odd $L$ to the ``subsystem'' $A$
	in the left plot and to the ``environment'' $B$ in the right plot.}
	\label{fig:EEPlot}
\end{figure}

For our purposes, we define the right-hand half of our system as the subsystem $A$ and the other half as the
``environment'' $B$.  We initialize the system in the state in which the leftmost spin is up and the rest down,
i.e., $\ket{\psi(t=0)}=\ket{\uparrow\downarrow\cdots\downarrow}$.  We are interested in the disorder average of
this quantity after a long evolution time, which we take to be $J_0t=100$.  In general, all of these calculations
must be done numerically, but we can obtain the per-realization entanglement entropy as a function of time
analytically in the $L=2$ case.  We find that it is
\begin{equation}
S_E=-x\ln{x}-(1-x)\ln(1-x),
\end{equation}
where
\begin{equation}
x=\frac{4J^2}{(\Delta h)^2+4J^2}\sin^2[(J+\sqrt{(\Delta h)^2+4J^2})t].
\end{equation}
We use $20,000$ realizations of disorder in this calculation.  For odd $L$, there is ambiguity as to whether the
middle spin is assigned to the subsystem $A$ or the ``environment'' $B$; we perform this calculation for both cases.
We present our results in Fig.\ \ref{fig:EEPlot}.  While the numerical values obtained for odd $L$ do depend on whether
we assign the middle spin to the subsystem $A$ or ``environment'' $B$, we find no change in the overall trend.
We note that the entanglement entropy is much larger, and increases with chain length, for $\sigma_h/J_0=0.1$
than it is for $\sigma_h/J_0=10$, for which value the entanglement entropy decreases slightly with length.
This again suggests that the system is in a delocalized ``phase'' for $\sigma_h/J_0=0.1$ and in an MBL
``phase'' for $\sigma_h/J_0=10$.

\begin{figure}
	\centering
		\includegraphics[width=0.49\columnwidth]{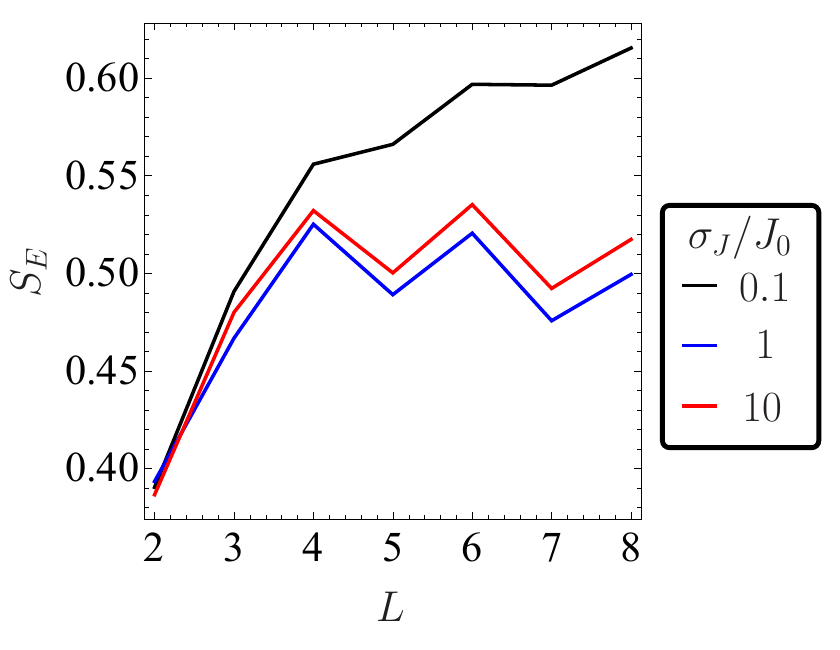}
		\includegraphics[width=0.49\columnwidth]{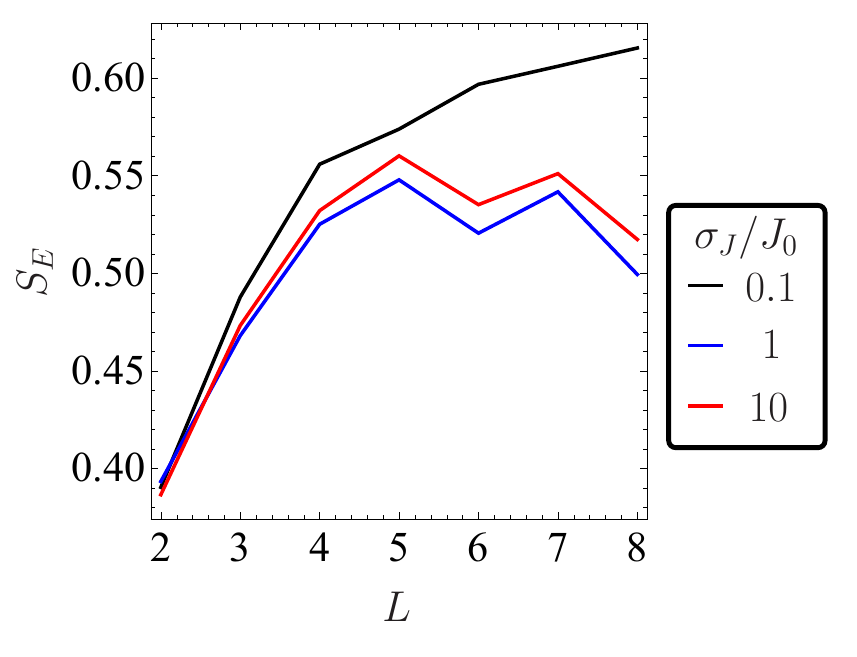}
	\caption{Plot of the entanglement entropy $S_E$ of the spin chain, with one half of the chain designated
	as the subsystem and the other as the ``environment'' as a function of spin chain length for $\sigma_J/J_0=0.1$,
	$1$, and $10$, at $J_0t=100$ and $\sigma_h/J_0=0.1$.  We assign the middle spin for odd $L$ to the ``subsystem'' $A$
	in the left plot and to the ``environment'' $B$ in the right plot.}
	\label{fig:EEPlot_varysJ}
\end{figure}

We repeat this calculation, this time fixing $\sigma_h$ and varying $\sigma_J$; we plot these results in
Fig.\ \ref{fig:EEPlot_varysJ}.  In this case, we again see no clear signs of localization; while $S_E$ plateaus
for longer chains, it does not change much when $\sigma_J$ is increased.
Again, as for Fig.\ \ref{fig:rsAllLengthsPlot_varysJ_1SUSSOnly}, this situation is specific to semiconductor spin qubits
where charge noise could enhance interaction effects in the system instead of only enhancing disorder effects.

\section{Spin-spin correlations} \label{sec:SpSpCorr}
\begin{figure*}
	\centering
		\includegraphics[width=0.32\textwidth]{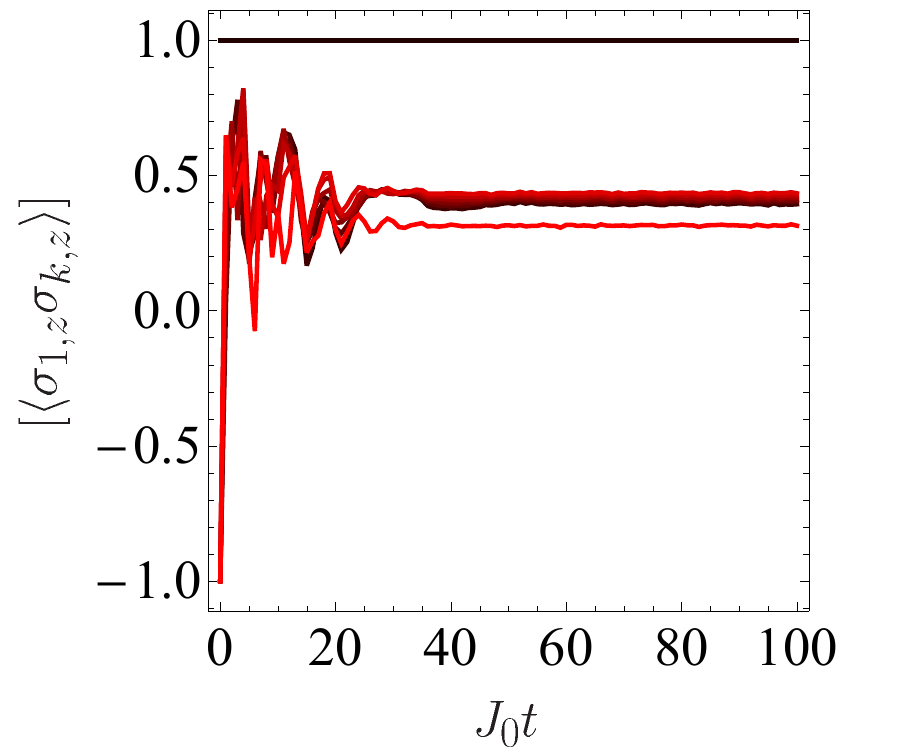}
		\includegraphics[width=0.32\textwidth]{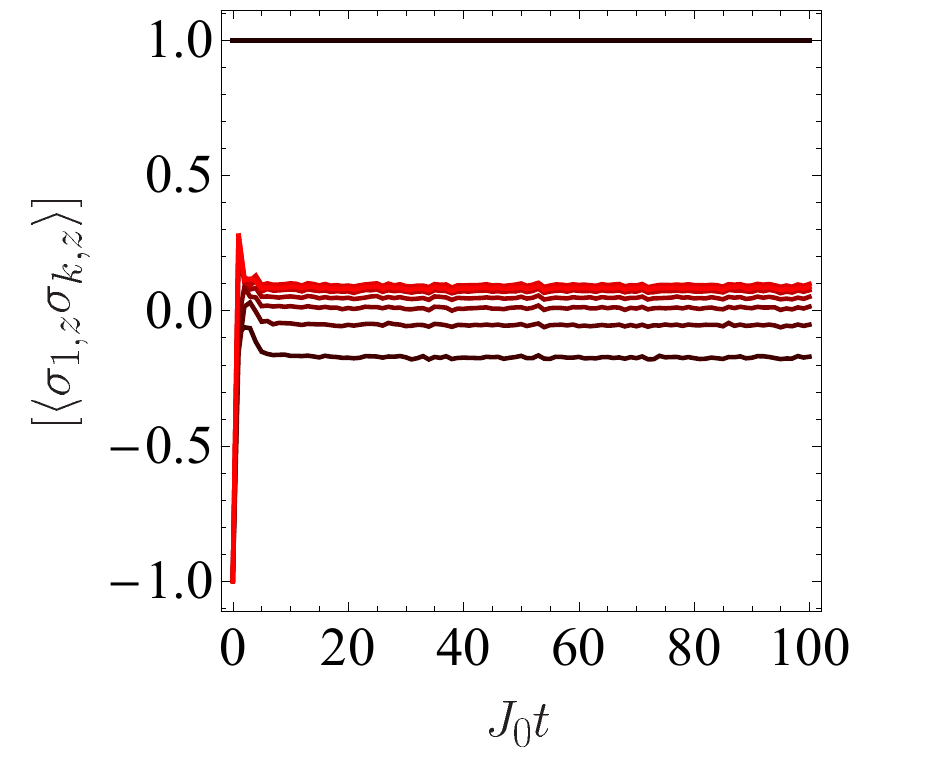}
		\includegraphics[width=0.32\textwidth]{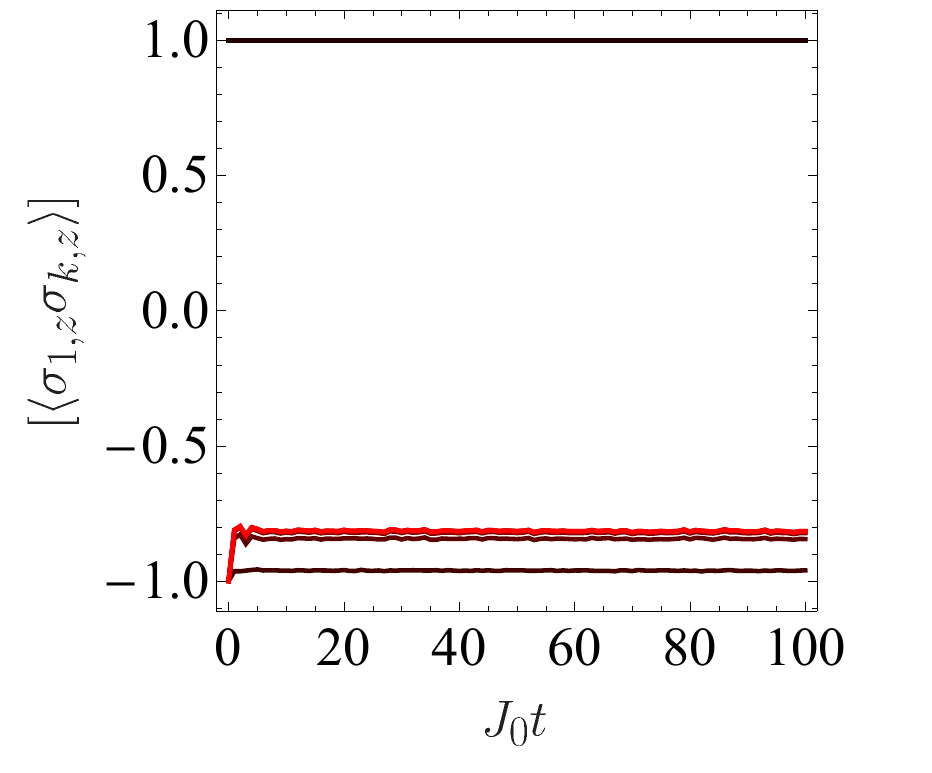}
	\caption{Disorder-averaged spin-spin correlation function $[\braket{\sigma_{1,z}\sigma_{k,z}}](t)$ for $L=8$ as a
	function of time for $\sigma_h/J_0=0.1$ (left), $1$ (center), and $10$ (right).  In each plot, $\sigma_J/J_0=0.1$
	and colors from black to red represent $k=1$ to $8$.}
	\label{fig:SpinCorr}
\end{figure*}

We now look at spin-spin correlations, which are experimentally relevant in spin qubit systems in contrast to
level spacing statistics and entanglement entropy considered above, in the sense that one can easily measure the
spin-spin correlation functions for a spin qubit system, but not level spacing statistics and the entanglement
entropy.  We focus specifically on two-point equal-time correlations,
$\braket{\sigma_{j,z}\sigma_{k,z}}(t)=\bra{\psi(t)}\sigma_{j,z}\sigma_{k,z}\ket{\psi(t)}$, and once again initialize the system
so that the left spin is up and the rest down, $\ket{\psi(t=0)}=\ket{\uparrow\downarrow\cdots\downarrow}$.
We predict that, in the delocalized ``phase'', the correlation functions should change significantly with time,
while, in the MBL ``phase'', there should be little change.  For reference, in the initial state, all other
spins should be anticorrelated with the leftmost spin, i.e., $\braket{\sigma_{1,z}\sigma_{k,z}}(t=0)=-1$ for any
$k\neq 1$.  We numerically calculate the disorder-averaged spin-spin correlation functions $[\braket{\sigma_{j,z}\sigma_{k,z}}](t)$
using $20,000$ realizations of disorder and present results for $L=8$ as a representative case in Fig.\
\ref{fig:SpinCorr}; the results for other values of $L\geq 3$ are qualitatively similar.  In the $L=2$ case,
the two spins remain perfectly anticorrelated for all times $t$ regardless of disorder; we may see this from
the fact that the Hamiltonian conserves total $S_z$, and thus, because we start purely in the $S_z=0$ sector, we
remain there for all times $t$.  This $L=2$ system is a trivial case of symmetry induced MBL, where the spins
have no freedom to thermalize no matter how strong (weak) the interaction (disorder) is.
We see that, for $\sigma_h/J_0=0.1$, the correlation function becomes positive for all positions $j$ after
a long time, indicating that all of the spins are significantly affected by interactions, direct or indirect,
with the leftmost spin.  However, if $\sigma_h/J_0=10$, then we find that the correlation functions deviate
little from their initial values, even at long times, indicating little influence from the leftmost spin.
Thus, for $\sigma_h/J_0=0.1$ ($10$), the $L=8$ system is delocalized (many-body-localized).

\begin{figure}
	\centering
		\includegraphics[width=\columnwidth]{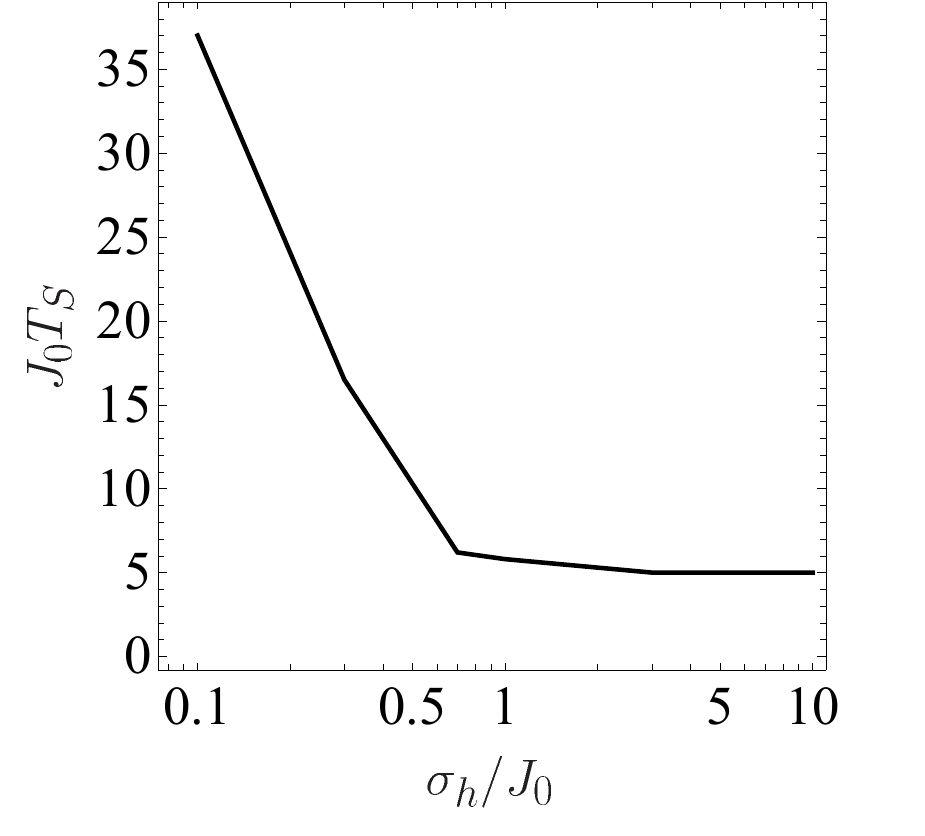}
	\caption{Plot of saturation time $T_S$ as a function of $\sigma_h$ for $L=8$.}
	\label{fig:TSPlot}
\end{figure}

We also notice that the spin-spin correlation functions achieve their long-time steady-state values over a
time scale, which we will denote by $T_S$, that depends on the value of $\sigma_h/J_0$.  More specifically, we
note that, for $\sigma_h/J_0=0.1$ and $L=8$, the correlation functions saturate at a time $J_0 t\approx 37$,
while, for $\sigma_h/J_0=10$ and the same $L$, they saturate at $J_0 t\approx 5$.  We plot this saturation
time $T_S$ as a function of $\sigma_h$ for $L=8$ in Fig.\ \ref{fig:TSPlot}.  We note a sharp drop from
$\sigma_h/J_0=0.1$ to $\sigma_h/J_0=0.7$, above which the plot remains mostly flat.

\begin{figure*}
	\centering
		\includegraphics[width=0.32\textwidth]{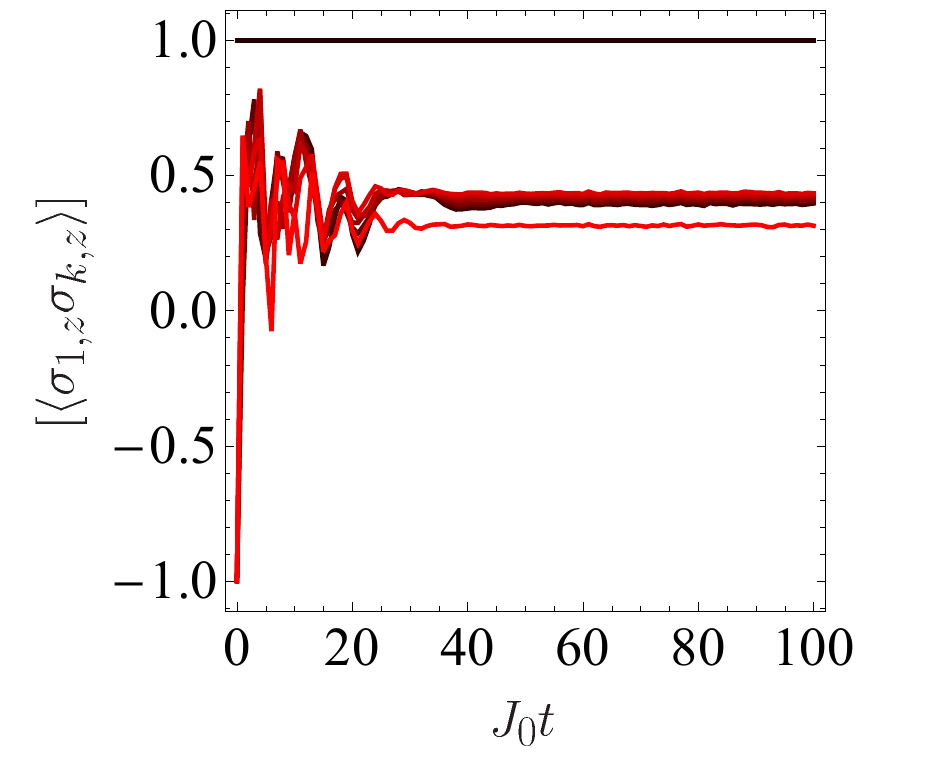}
		\includegraphics[width=0.32\textwidth]{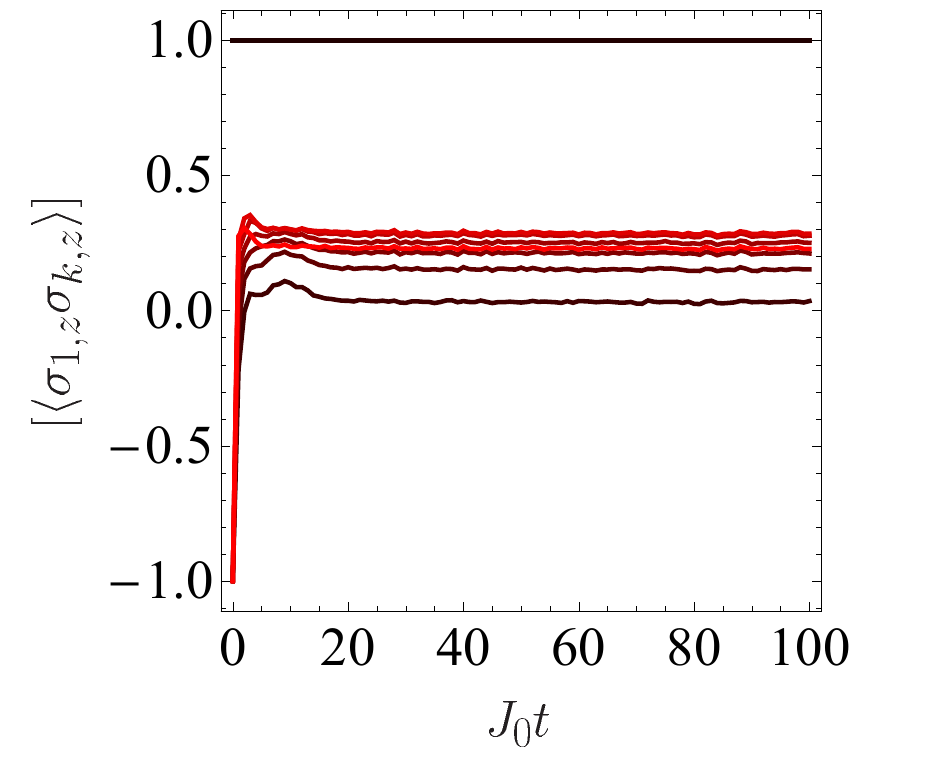}
		\includegraphics[width=0.32\textwidth]{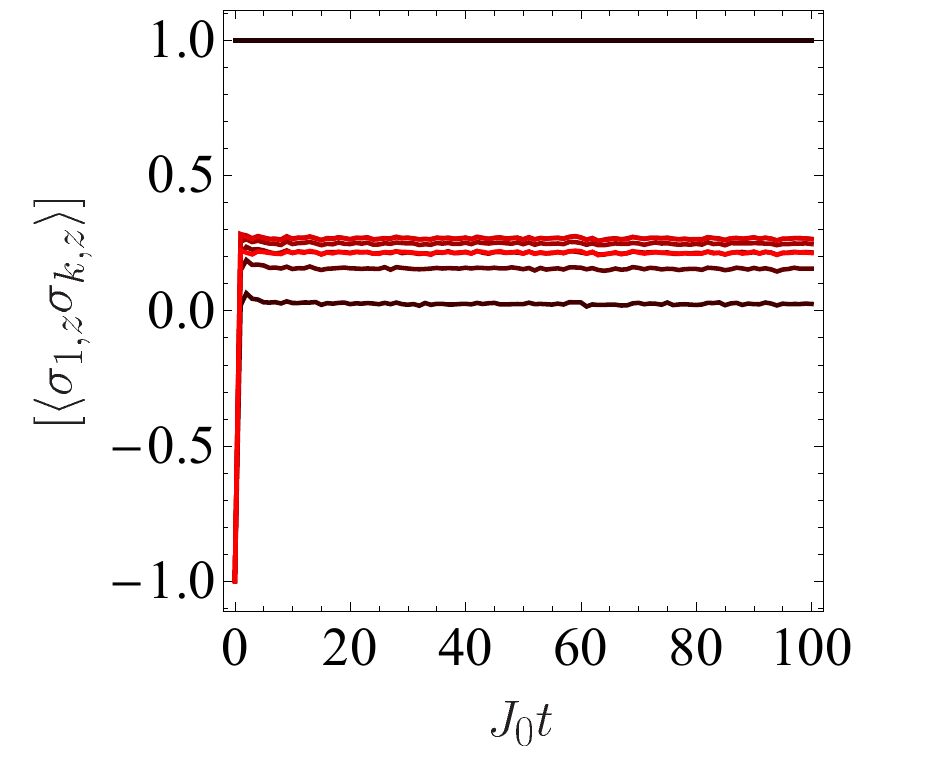}
	\caption{Disorder-averaged spin-spin correlation function $[\braket{\sigma_{1,z}\sigma_{k,z}}](t)$ for $L=8$ as a
	function of time for $\sigma_J/J_0=0.1$ (left), $1$ (center), and $10$ (right).  In each plot, $\sigma_h/J_0=0.1$
	and colors from black to red represent $k=1$ to $8$.}
	\label{fig:SpinCorr_varysJ}
\end{figure*}

In the case in which we fix $\sigma_h$ and vary $\sigma_J$, however, it is more difficult to identify a
single time $T_S$ at which oscillations cease, as different spins stop oscillating at different times.
We plot our results in this case in Fig.\ \ref{fig:SpinCorr_varysJ}.
We also note that there is hardly any signature for localization for any value of $\sigma_J$ in this case.  This
is not surprising, as we found that, from the level spacing ratios, the system never tends (see Fig.\ 2) toward clear
localization for any value of $\sigma_J$ because of the effective enhancement of interaction (i.e., exchange coupling)
when the charge disorder increases.  Thus, Figs.\ \ref{fig:rsAllLengthsPlot_varysJ_1SUSSOnly} (level statistics), \ref{fig:EEPlot_varysJ}
(entanglement entropy), and \ref{fig:SpinCorr_varysJ} (spin correlator) are consistent with each other establishing
that increasing charge disorder in the system may not lead to MBL in the spin qubit systems in contrast to increasing
the field disorder (Figs.\ \ref{fig:rsAllLengthsPlot_1SUSSOnly}, \ref{fig:EEPlot}, and \ref{fig:SpinCorr}) which would induce
MBL with increasing disorder.  This inconsistency between charge and field disorder is intrinsic to spin qubits and
is an important finding of our work.

Spin correlation results for other values of $L$ ($4$--$7$) are similar to those shown in Figs.\ \ref{fig:SpinCorr} and
\ref{fig:SpinCorr_varysJ}, and are therefore not shown here.

\section{Conclusion} \label{sec:Conc}
We propose an experimentally relevant metric by which one may determine whether or not a semiconductor
spin qubit system is in an MBL or a delocalized ``phase'', the (dimensionless) disorder-averaged spin-spin
correlation function $[\braket{\sigma_{j,z}\sigma_{k,z}}](t)$.  We propose this metric in particular
because it is particularly well suited for use in detecting MBL effects in semiconductor-based electron
spin qubit systems currently being used for quantum computation experiments due to the fact that the
individual spins may be measured and, from these measurements, the spin-spin correlation functions can
be obtained.  We calculated the spin-spin correlation function along with the long time entanglement entropy,
the latter a standard metric for establishing the presence or absence of localization,
for a finite chain of Heisenberg-coupled spins of lengths $L=2$ to $8$ in the presence of field and
charge noise as arising in semiconductor spin qubit systems, as well as the average level spacing ratio
for systems of lengths $L=3$ to $10$.  In determining the entanglement entropy
and spin-spin correlation functions, we assumed that the system was initialized so that the leftmost
spin was up and the rest down, i.e., $\ket{\psi(0)}=\ket{\uparrow\downarrow\cdots\downarrow}$.
We determined these quantities as functions of $\sigma_h$ for fixed $\sigma_J$, and vice versa.  We
demonstrated that the behavior of the spin-spin correlation function works well as a metric for identifying
the delocalized and localized ``phases'', providing results consistent with those obtained from the more
standard level statistics and entanglement entropy studies---we find that the spin-spin correlator deviates
significantly from its initial value if the system is in the delocalized ``phase'' according to the other metrics,
while it stays close to its initial value in the localized ``phase''.  We also note that, when varying $\sigma_h$
with $\sigma_J$ held constant, the spin-spin correlation function $[\braket{\sigma_{1,z}\sigma_{k,z}}](t)$,
independent of $k$, achieves its steady-state value over a time scale $T_S$.  We determine this time and plot
it as a function of $\sigma_h$, and find that it decreases with increasing $\sigma_h$ in the delocalized ``phase''
and remains roughly constant in the localized ``phase''.  These findings do not change if the chosen initial
state is different from the specific choice we made in our work, and our choice of one spin up and the rest
down is inspired entirely by its convenient experimental realization.  Since our results only depend on the
dimensionless quantities, $\sigma_h/J_0$ and $\sigma_J/J_0$, experiments intended to observe the effects that
we predict can be performed by changing the value of $J_0$, the intended exchange coupling between two nearest-neighbor
spins.  This is in fact the only parameter that can be reliably controlled in a semiconductor-based electron
spin qubit system.  By doing so, one can sweep over a range of values of $\sigma_h/J_0$, with lower values of
$J_0$ resulting in the localized ``phase'' and larger values of $J_0$ yielding the delocalized ``phase''.  In
fact, the experimental system considered in Ref.\ \onlinecite{MartinsPRL2016} is capable of measurements in
both the localized and delocalized regimes---the value of $\sigma_h=23\,\text{MHz}$, and values of $J$ ranging
from $40\,\text{MHz}$ to $700\,\text{MHz}$ are possible.  While $\sigma_J$ may vary as a function of $J_0$,
this variation will likely have little effect on whether the system is localized or delocalized, as implied
by our results for the level spacing ratio $[r^{(n)}_\alpha]$ as a function of $\sigma_J$ for levels within
the $S_z=1-\tfrac{1}{2}L$ subspace given in Fig.\ \ref{fig:rsAllLengthsPlot_varysJ_1SUSSOnly}.  We must also
emphasize that the results that we have presented do not make any reference to how $\sigma_J$ varies as a
function of $J_0$.

The advantage of the spin-spin correlation function over the other two metrics considered here is that
it is particularly well suited to the study of MBL effects in existing experimental semiconductor-based
electron spin qubit systems.  While much progress has been made over the last several years in spin qubit
systems, single- and two-qubit gates in these qubits have not yet achieved the fidelities (at least $99.9\%$)
required for error-correcting techniques to be effective.  In addition, in contrast to the superconducting
qubits and ion trap qubits, the number of operational qubits at this stage  in the semiconductor system
is rather low ($L=2$--$8$ at most).  It is therefore premature to discuss quantum computing or quantum
supremacy using semiconductor qubit platforms, and the only concrete achievements so far have been ground
state many body simulations with $2$--$4$ qubits\cite{DehollainNature2020,WangPRB2019}.
However, such systems can be used to realize MBL effects, as the very effects (i.e., noise) that are detrimental
to the use of these qubits for quantum computing make them well suited to the study of such effects.  Our
proposal for studying MBL using spin correlators will be a significant achievement for semiconductor qubits
since the study would be dynamical with disorder and noise along with the exchange coupling playing key roles.
A chain of exchange-coupled electron spins subject to Overhauser (magnetic) and electronic (exchange) noise
is analogous (in fact mathematically equivalent) to a one-dimensional tight-binding chain model
with a random on site potential and randomized tight-binding hopping terms.  For this purpose, one does
not need precise control over the spins in order to perform these experiments; one only needs to measure
the system after its natural evolution.  The spin-spin correlation function may then be obtained simply
from measurements of the individual spins' states as a function of time.  The only source of error in such
an experiment would come from the preparation of the initial state and readout of the final state.  A delocalized
``phase'' is then characterized by significant change in the correlation function as a function of time,
in our example a long-time value differing greatly from $-1$ for two different spins, while a localized
``phase'' would, again in our example, be characterized by the correlation functions staying negative with
absolute values of order unity (say, not much less than $0.5$).  While we presented this measure in
the context of a spin chain, there is nothing that, in theory, limits the use of this measure to such a
system.  In fact, one could easily adapt this measure to charge qubits as well.

Finally, we point out an interesting finding of our work, which is specific to exchange-coupled spin qubits
in semiconductor quantum dots, where the Heisenberg coupling is of a fixed sign (always positive, at least at
typical magnetic fields used in experiments).  This makes the Overhauser field noise and electronic charge noise
have qualitatively different effects on the many body localization dynamics of the system.  While increasing
the field noise necessarily leads to many body localization in a given system, increasing the charge noise
generically does not.  In fact, we find no clear signs of localization as we increase charge noise.
This happens because the charge disorder effectively enhances the Heisenberg exchange coupling in the system,
and therefore, increasing charge noise often is equivalent to increasing interaction in the system.  This
peculiar effect of charge noise not hindering delocalization in spin qubits could be directly experimentally
tested in existing spin qubit systems.  We note here that our calculations do not include valley effects, making
them more relevant to GaAs systems than to Si systems.  Not only that, but Si systems would have very little,
if any, Overhauser field noise due to isotopic purification.  While these issues are important, they are beyond
the scope of the present work.

\acknowledgments
This work was funded by the Laboratory for Physical Sciences.

\appendix

\section{Average level spacing ratio for large system sizes} \label{app:AvgLSR_LargeSystems}
\begin{figure}
	\centering
		\includegraphics[width=\columnwidth]{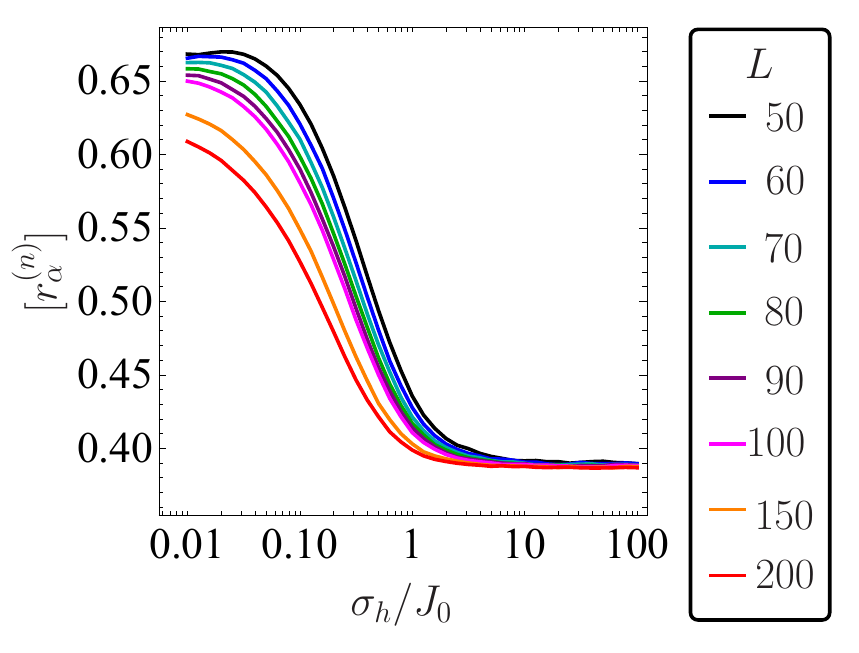}
		\caption{Disorder-averaged level spacing ratios $[r^{(n)}_\alpha]$ as a function of $\sigma_h/J_0$ for $L=50$
		to $200$ and $\sigma_J/J_0=0.1$, calculated only using levels in the $S_z=1-\tfrac{1}{2}L$ subspace.}
	\label{fig:rsAllLengthsPlot_1SUSSOnly_LargeSystems}
\end{figure}

\begin{figure}
	\centering
		\includegraphics[width=\columnwidth]{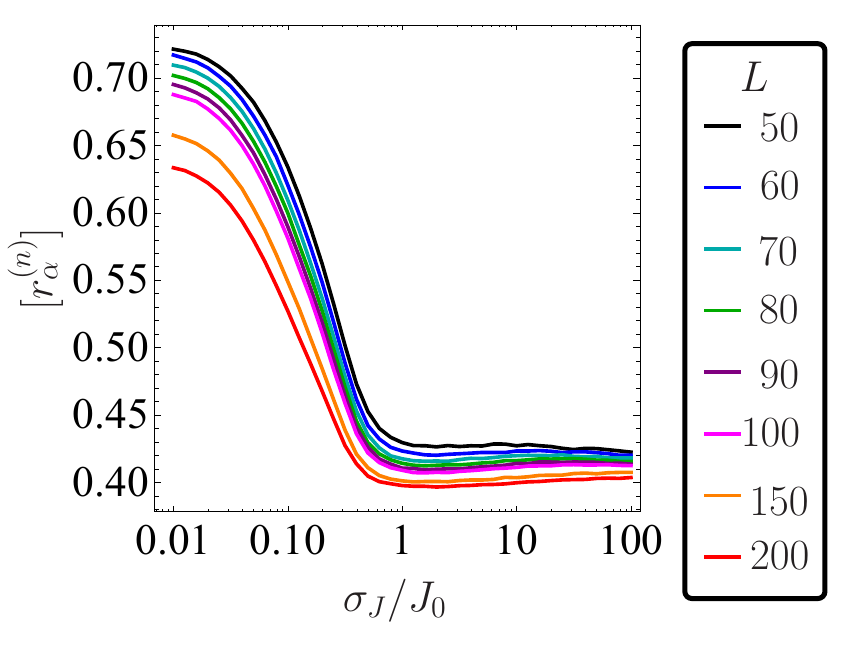}
		\caption{Disorder-averaged level spacing ratios $[r^{(n)}_\alpha]$ as a function of $\sigma_J/J_0$ for $L=50$
		to $200$ and $\sigma_h/J_0=0.1$, calculated only using levels in the $S_z=1-\tfrac{1}{2}L$ subspace.}
	\label{fig:rsAllLengthsPlot_varysJ_1SUSSOnly_LargeSystems}
\end{figure}

We present here our results for the average level spacing ratio for large systems ($L=50$--$200$), both as a
function of $\sigma_h$ for fixed $\sigma_J=0.1J_0$ and as a function of $\sigma_J$ for fixed $\sigma_h=0.1J_0$,
which are shown in Figs.\ \ref{fig:rsAllLengthsPlot_1SUSSOnly_LargeSystems} and \ref{fig:rsAllLengthsPlot_varysJ_1SUSSOnly_LargeSystems}.
In generating both plots, we used $10,000$ realizations of disorder.
Not surprisingly, the tendencies towards delocalization for small $\sigma_h$ and localization for large $\sigma_h$
are enhanced at larger system sizes, with the small $\sigma_h$ limit tending more toward the expected GOE value of
$0.53$ as $L$ is increased.  We also see that a clear tendency toward the expected behavior as a function of $\sigma_J$
of delocalization for small $\sigma_J$ and localization for large $\sigma_J$ appears for large $L$.

\section{Average level spacing ratio calculated using all energy levels} \label{app:AvgLSR_AllEnLevels}
\begin{figure}
	\centering
		\includegraphics[width=\columnwidth]{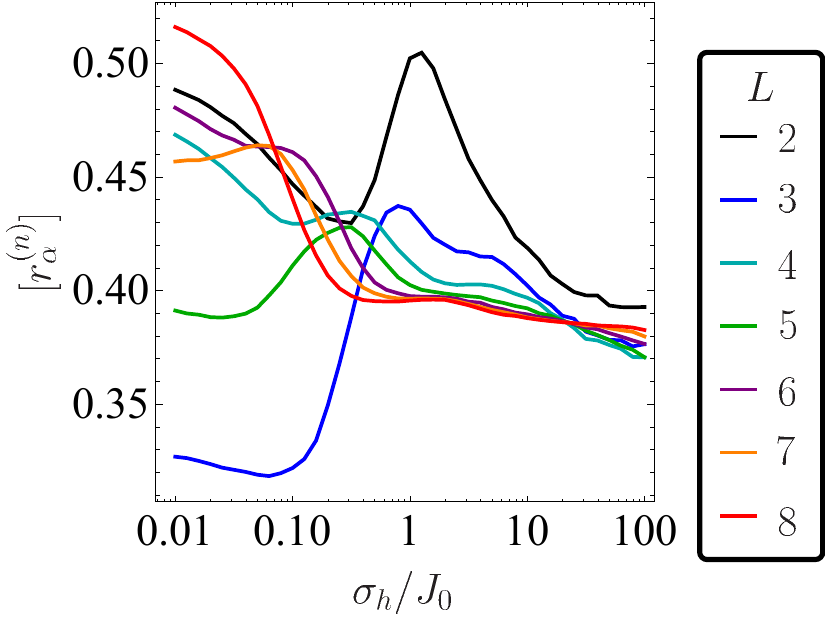}
		\caption{Disorder-averaged level spacing ratios $[r^{(n)}_\alpha]$ as a function of $\sigma_h/J_0$ for $L=2$
		to $8$ and $\sigma_J/J_0=0.1$, calculated using all energy levels.}
	\label{fig:rsAllLengthsPlot}
\end{figure}

\begin{figure}
	\centering
		\includegraphics[width=\columnwidth]{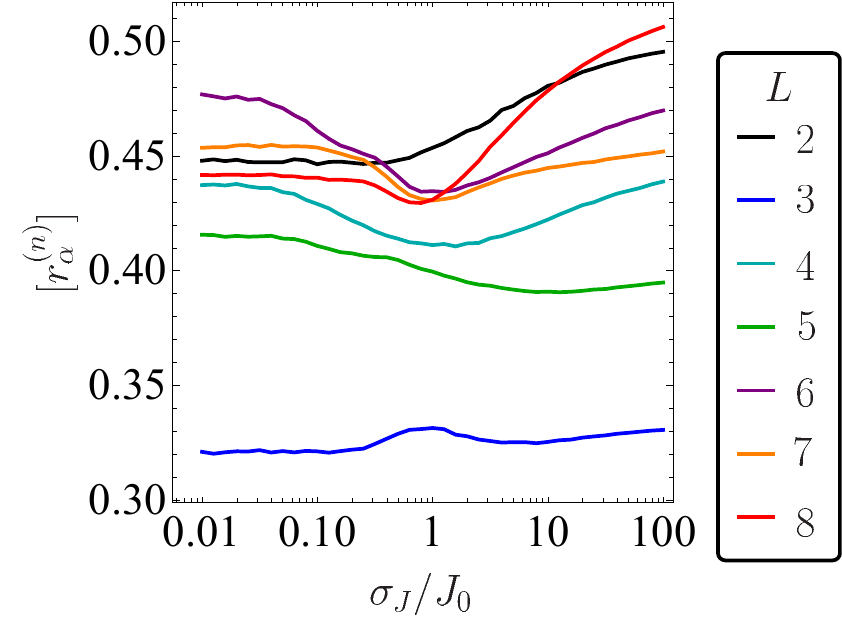}
		\caption{Disorder-averaged level spacing ratios $[r^{(n)}_\alpha]$ as a function of $\sigma_J/J_0$ for $L=2$
		to $8$ and $\sigma_h/J_0=0.1$, calculated using all energy levels.}
	\label{fig:rsAllLengthsPlot_varysJ}
\end{figure}

Here, we present our results for the average level spacing for $L=2$--$8$, calculated using all energy levels,
not just those within the $S_z=1-\tfrac{1}{2}L$ subspace.  We show these results in Figs.\ \ref{fig:rsAllLengthsPlot}
and \ref{fig:rsAllLengthsPlot_varysJ}.  We see that these results imply the same qualitative conclusions that
one can draw from the $S_z=1-\tfrac{1}{2}L$ subspace.  We note that, unlike with the $S_z=1-\tfrac{1}{2}L$
results, there is no overshoot of the GOE value of $0.53$.

\end{document}